\documentclass[twocolumn]{aastex631}
\newcommand{\mstar}{ M_{\star}}
\newcommand{\sfr}{\rm SFR}

\def\beq{\begin{equation}}
\def\eeq{\end{equation}}
\def\beqn{\begin{eqnarray}}
\def\eeqn{\end{eqnarray}}

\newcommand{\OIII}{[O\,{\sc iii}]}
\newcommand{\Ha}{H$\alpha$}
\newcommand{\Hb}{H$\beta$}
\newcommand{\NII}{[N\,{\sc ii}]}

\newcommand{\HI}{H{\sc i}}
\newcommand{\Oabundance}{12+\log({\rm O/H})}

\usepackage{ulem}
\usepackage{amsmath}
\usepackage{multirow}
\usepackage{hyperref}
\usepackage{array}
\usepackage{makecell}

\begin{document}

\title{
No Metallicity Preference in Fast Radio Burst Host Galaxies
}

\correspondingauthor{Shotaro Yamasaki}
\email{shotaro.s.yamasaki@gmail.com}

\author[0000-0002-1688-8708]{Shotaro Yamasaki}
\affiliation{Department of Physics, National Chung Hsing University, 145 Xingda Rd., South Dist., Taichung 40227, Taiwan}

\author[0000-0001-7228-1428]{Tetsuya Hashimoto}
\affiliation{Department of Physics, National Chung Hsing University, 145 Xingda Rd., South Dist., Taichung 40227, Taiwan}

\author[0000-0002-3801-434X]
{Haruka Kusakabe}
\affiliation{Department of General Systems Studies, Graduate School of Arts and Sciences, The University of Tokyo, Meguro, Tokyo, 153-8902, Japan}

\author[0000-0002-6821-8669]{Tomotsugu Goto}
\affiliation{Department of Physics, National Tsing Hua University, 101, Section 2. Kuang-Fu Road, Hsinchu, 30013, Taiwan}
\affiliation{Institute of Astronomy, National Tsing Hua University, 101, Section 2. Kuang-Fu Road, Hsinchu, 30013, Taiwan}

\begin{abstract}

Fast radio bursts (FRBs) are millisecond-duration extragalactic radio transients of unknown origin, and studying their host galaxies could offer clues to constrain progenitor models. Among host properties, gas-phase metallicity is a key factor influencing stellar evolution and transient productions. We analyze the largest uniformly selected sample of FRB host galaxies, measuring oxygen abundances ($12+\log(\mathrm{O/H}) = 8.04$–$8.85$) for 31 hosts at redshifts $z = 0.04$–$0.98$, using consistent emission-line diagnostics. Using a volume-limited subsample, we compare the distributions of stellar mass, star formation rate (SFR), and metallicity to a control sample of star-forming galaxies selected by the same criteria. We find that FRB host galaxies span a wide metallicity range and are broadly consistent with the SFR-weighted mass–metallicity relation of star-forming galaxies. We find no clear lower metallicity bound, suggesting that FRB progenitors can form in any metallicity environment through channels largely insensitive to metal abundance. Encouragingly, this implies FRBs can arise even in low-metallicity, high-redshift galaxies, supporting their potential as probes of matter distribution across cosmic time.
Additionally, we find marginal ($\sim$2$\sigma$) evidence for a $-0.09 \pm 0.04$ dex metallicity offset from the fundamental metallicity relation. Despite model uncertainties, if real, this offset likely reflects suppressed SFRs at fixed mass and metallicity rather than metal deficiency. 
Similar offsets are observed in local post-merger galaxies and may reflect a post-starburst phase following galaxy interactions. Such systems may host FRB progenitors formed during the starburst that produce FRBs after a 100–500 Myr delay, broadly consistent with observed delay-time distributions, although further data are needed to confirm this interpretation.
\end{abstract}

\keywords{Radio transient sources (2008) --- Metallicity (1031) --- Galaxy evolution (594)}

\section{Introduction}
\label{section:intro}

Fast radio bursts (FRBs) are millisecond-duration radio transients of extragalactic origin \citep{lorimer07,petroff22}. While their progenitors remain unknown, magnetars—strongly magnetized neutron stars—are widely discussed as leading candidates \citep[see][for review]{zhang20}.
Thanks to the growing number ($\sim$100) of identified FRB host galaxies \citep[e.g.,][]{bhandari22,gordon23,law24,bhardwaj24,sharma24,shannon25,chimekko25,gordon25}, statistical studies of their environments have now become possible, offering new insights into their potential formation pathways.

Among the various properties of host galaxies, metallicity fundamentally shapes stellar evolution and has been recognized as a critical parameter for constraining the progenitors of transient phenomena, including Type Ia supernovae (SNe) \citep[e.g.,][]{pan14}, Type II SNe \citep[e.g.,][]{stoll13,taggart21}, Superluminous SNe \citep{perley16,chen17} and gamma-ray bursts (GRBs) \citep[e.g.,][]{levesque10,perley16,niino16}.   For instance, studies of long GRBs revealed a preference for low-metallicity host galaxies \citep{stanek06}, based primarily on gas-phase metallicity measurements of their star-forming regions, leading to the development of the collapsar model, where metal-poor progenitor stars can retain enough angular momentum to launch relativistic jets \citep{woosley93}. By analogy, measuring the gas-phase metallicity of FRB host galaxies would provide a promising avenue to test whether metallicity plays a role in shaping FRB progenitor channels—potentially distinguishing, for example, between magnetars formed via core-collapse SNe \citep[e.g.,][]{metzger17,kashiyama17}, compact object mergers \citep[e.g.,][]{yamasaki18,margalit19}, dynamical interactions in globular clusters \citep[e.g.,][]{kremer21,li24}, magnetars born in binary systems \citep[][]{zhang17,ioka20,wada21,Zhang2025}, or other alternative pathways (see \citealt{zhang20} for a review).

To date, most FRB host studies have focused on the distributions of stellar mass and star formation rate (SFR), either individually or jointly \citep[e.g.,][]{bhandari20,heintz20,bochenek21,bhandari22,gordon23,law24,sharma24,loudas25,horowicz25}. Recent work by \citet{sharma24} suggested that the stellar mass distribution of FRB hosts shows a lower bound, which may hint at a metallicity threshold currently under debate \citep[e.g.,][]{horowicz25,august25}. However, it is important to note that these studies examined only the distributions of stellar mass and/or SFR, without directly measuring or analyzing metallicity itself in a sufficiently large sample.

In this work, we present the first large-sample study of gas-phase metallicity in FRB host galaxies, based on strong emission-line measurements available in the literature. Unlike stellar metallicity, which reflects the past integrated history, gas-phase metallicity traces the present conditions of star-forming regions where massive stars form. This makes it a more direct probe of the environments relevant to FRB progenitors.
By investigating metallicity jointly with stellar mass and SFR, we aim to directly test whether FRB progenitors prefer specific metallicity environments. Furthermore, as a methodological improvement over previous FRB host studies, we construct a volume-limited sample—rather than a purely flux-limited one—to reduce optical selection biases. By consistently applying this volume-limited selection to both the observed FRB host sample and the comparison galaxy sample, we can robustly compare their distributions and better constrain the environments of FRB progenitors.

This {\it Letter} is organized as follows. In \S \ref{s:data}, we describe the construction of the selected FRB host sample and the estimation of their metallicities. \S \ref{s:model} presents the modeling of the comparison galaxy sample.
Our results are shown in \S \ref{s:result}, followed by a discussion of their implications in \S \ref{s:discussion}.
Finally, we summarize our findings and conclude in \S \ref{s:conclusion}. 
Throughout, we assume a standard $\Lambda$CDM cosmology ($\Omega_{\rm m} = 0.3$, $\Omega_{\Lambda} = 0.7$, $h = 0.7$) and refer to gas-phase metallicity simply as ``metallicity'' unless stated otherwise. 

\section{FRB Host Galaxy Sample}
\label{s:data}

\subsection{Parent Host Sample}

In this study, we require measurements of stellar mass, star formation rate (SFR), and an independent gas-phase metallicity estimate for FRB host galaxies.
We base our analysis on the spectroscopically identified sample of FRB host galaxies compiled by \citet{sharma24}, which comprises 23 hosts compiled by  \citet{gordon23}—18 of which were localized by ASKAP and the remaining 5 by other facilities (see their Table 1)— and 26 hosts from DSA-110  \citep{sharma24}, for a total of $N=49$ host galaxies spanning a redshift range of $z=0.011$--$0.975$ (``Step 0'' in Table \ref{tab:selection}). For stellar mass and SFR, we adopt values derived using the \texttt{Prospector} code \citep{leja22} with non-parametric star formation histories (SFHs), as provided by \citet{sharma24} and \citet{gordon23}.

To estimate gas-phase metallicity, we use measurements of the \OIII$\lambda$5007, \Hb, \Ha, and \NII$\lambda$6584 emission lines (see \S~\ref{ss:metal}). We exclude FRBs 20190711A, 20210117A, 20180916B, and 20211212A because either (i) at least one of these four lines lacks a usable flux measurement (i.e., is undetected without an upper limit), or (ii) both lines in at least one of the two pairs—\OIII$\lambda$5007 and \Hb\ or \NII$\lambda$6584 and \Ha—are detected only as upper limits, preventing a constraint on the corresponding line ratio. After these exclusions, we retain $N=45$ host galaxies, which we define as our {\it parent} sample (after ``Step 1'' in Table~\ref{tab:selection}; full sample listed on Table~\ref{tab:frb_host_data} in Appendix~\ref{s:fullsample}).

We do not distinguish between repeaters and non-repeaters in this work, as our parent sample is dominated by apparent non-repeaters 
(41 out of 45), making statistical comparisons challenging. Observationally, repeater and non-repeater signals show distinct differences in pulse width and bandwidth \citep{pleunis21}. Nevertheless, recent studies suggest that most apparent non-repeaters may actually be repeaters \citep[e.g.,][]{james23,yamasaki24,sun25}, and find no significant differences in host properties between the two populations \citep{gordon23,sharma24}

\begin{table*}
\centering
\caption{Sample selection summary for FRB host galaxies. The number of hosts remaining after each selection step is shown for two different samples used in this work (see \S \ref{s:data} for details). The initial sample (at Step 0) was compiled from \citet{eftekhari23,gordon23,sharma24}. Our {\it parent} sample is defined as galaxies that pass selection Step 1. For the \textit{No-volume-limit} sample, Step 2 is skipped.}
\label{tab:selection}
%\begin{tabular}{l p{7cm} c c}
\begin{tabular}{l l c c}
%\begin{tabular}{@{} l >{\raggedright\arraybackslash}p{0.55\textwidth} c c @{}}
\hline\hline 
%Selection step & \makecell[l]{Satisfying the adopted redshift \\ and completeness cut; $0.01<z<0.48$ \\ $M_r<-17.5$} & 39 & -- \\
\multirow{2}{*}{Selection step}&\multirow{2}{*}{\makecell[l]{Selection criteria}}&\multicolumn{2}{c}{\textbf{Number of FRB hosts}}\\
&&\textit{Volume-limited}&\textit{No-volume-limit} 
\\[3pt]
\hline\\[-8pt]
(0) Initial sample &  \makecell[l]{Compiled from literature} & 49 & 49 
\\[8pt]
(1) Emission-line availability & \makecell[l]{Sufficient line detections to be plotted on the \\BPT diagram} & 45 & 45  
\\[8pt]
(2) Volume-limited selection & \makecell[l]{Satisfying the adopted redshift and completeness \\cut; $0.01 < z < 0.48$ \& M$_r < -17.5$} & 39 & -- 
\\[8pt]
(3) BPT-diagram selection & \makecell[l]{Classified as star-forming on the BPT diagram; \\AGN/LINER sources excluded} & 29 & 34
\\[8pt]
(4) Metallicity measurement & \makecell[l]{O3N2 can be determined; both O3 and N2 \\detected, or one is an upper limit and the other \\a lower limit} & 27 ($+$1 UL) & 31 ($+$2 UL/LL)
\\[9pt]
\hline
\end{tabular}
%\end{tabulary}
\end{table*}

\begin{figure*}
\centering
\includegraphics[width=17cm]{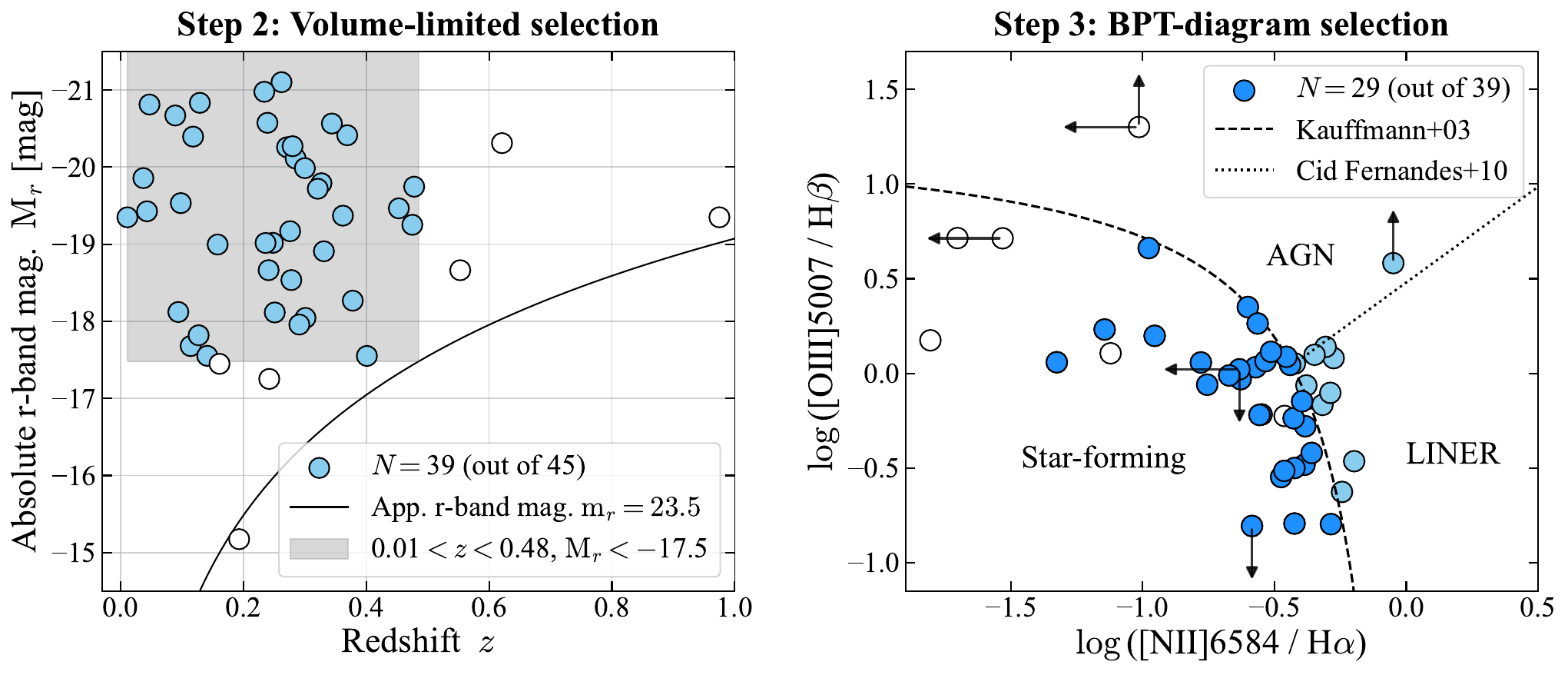}
\caption{Host galaxy selection following Steps 2 and 3 of the ``volume-limited'' sample described in Table \ref{tab:selection}.
Step 2 ({\it left}): Volume-limited selection in the redshift–absolute r-band magnitude plane. The solid curve shows the apparent magnitude limit of m$_r=23.5$, used as a reference to indicate the approximate depth of the host galaxy samples based on the completeness limit of the DSA sample in \citet{sharma24}. 
The gray shaded region indicates our selection window, optimized to include the maximum number of sources (light blue filled circles). Six sources (open circles) fall outside this window and are excluded.
Step 3 ({\it right}): BPT-diagram-based selection of star-forming galaxies. From the Step 1 sample, we select only those classified as SF galaxies based on emission line ratios to construct the final sample (dark blue filled circles). Black arrows indicate data points with upper or lower limits. The sample was compiled from \citet{eftekhari23,gordon23,sharma24}.}
\label{fig:selection}
\end{figure*}

\subsection{Volume-Limited Selection}
\label{ss:vl-selection}

We construct a volume-limited sample of FRB host galaxies for the first time in the FRB host galaxy study, rather than using a purely flux-limited sample. A flux-limited sample often used in previous FRB host galaxy studies \citep[e.g.,][]{sharma24,loudas25,horowicz25}, which includes all galaxies brighter than a given apparent magnitude (e.g., m$_r < 23.5$ mag), is straightforward to build and typically provides a larger number of objects. However, it inevitably suffers from selection biases: at higher redshifts, only intrinsically brighter (more massive and more metal-rich) galaxies are observable, while fainter galaxies drop out. This bias can artificially skew the inferred distributions of stellar mass, SFR, and gas-phase metallicity of FRB host galaxies.

To mitigate these biases, we define a volume-limited sample by selecting all host galaxies within a redshift range ($z < z_{\rm max}$) where the survey is complete down to a certain absolute magnitude threshold. Specifically, we determine $z_{\rm max}$ by ensuring that galaxies brighter than the absolute r-band magnitude M$_r$ (corresponding to the apparent magnitude limit of m$_r$ at $z_{\rm max}$) are fully included in our sample. This allows us to compare the properties of FRB host galaxies with the modeled galaxy population in an unbiased way\footnote{Ideally, radio selection biases should also be taken into account. However, because of the large uncertainties in the underlying FRB population, it is not currently possible to correct for these biases \citep{horowicz25}, and they are therefore not considered in this work. This omission is likely to have little impact unless there is a systematic correlation between FRB energy and host galaxy mass (or absolute magnitude), which to our knowledge has not been reported in the literature.}. While adopting a volume-limited sample necessarily reduces the number of FRB hosts in the analysis, it ensures that our statistical comparisons and interpretations reflect the true underlying population, rather than being dominated by observational selection effects.

The left panel of Figure \ref{fig:selection} shows the distribution of redshift versus absolute magnitude for the entire ($N=45$) parent host galaxy sample\footnote{In addition to the flux-limited bias corrected here, we note that host galaxy association itself can introduce a potential bias. 
Some FRBs lack secure host identifications (e.g., 16/42 in \citealt{sharma24}), and criteria such as requiring sufficiently high 
host spectrum S/N (as in \citealt{gordon23}) may preferentially select brighter galaxies. However, the galaxies missed in this way 
lie well below the magnitude limit shown in the left panel of Figure \ref{fig:selection}, so our results are unlikely to be significantly affected as long as the optical follow-up sensitivities 
are broadly comparable across surveys. 
}. We adopt an apparent magnitude limit of $m_r = 23.5$, corresponding to the completeness limit of the DSA sample in \citet{sharma24}, and use it as a reference to indicate the approximate depth of the combined host galaxy samples rather than a strict completeness boundary.
We adopt a selection criterion of $0.01 < z < 0.48$ and M$_r < -17.5$, chosen to maximize the number of selected hosts given the apparent magnitude limit of m$_r=23.5$. This results in a remaining sample of $N=39$ hosts out of $45$ (after Step~2 of ``volume-limited'' sample in Table \ref{tab:selection} and Figure \ref{fig:selection}).

\subsection{Star-Forming Host Selection with BPT Diagram}
\label{ss:bpt-selection}

In addition to applying the volume-limited selection (\S \ref{ss:vl-selection}), we further refine our FRB host galaxy sample by excluding galaxies classified as active galactic nuclei (AGNs) or low-ionization nuclear emission-line regions (LINERs).
There are two main motivations for this choice. First, the observed nebular emission lines in these systems cannot be explained solely by star formation, but instead require significant contributions from AGN activity. Second, estimates of SFR and gas-phase metallicity are considered robust and physically meaningful only for galaxies where star formation dominates the ionization, because standard metallicity diagnostics have been calibrated on such star-forming galaxies.

We adopt the Baldwin–Phillips–Terlevich (BPT) diagram \citep{bpt81}, shown in the right panel of Figure~\ref{fig:selection}, to classify our sample. The required \OIII$\lambda$5007, \Hb, \Ha, and \NII$\lambda$6584 line measurements of FRB hosts (after corrected for Galactic
extinction) are taken from the literature \citep{eftekhari23,sharma24}.\footnote{\citet{sharma24} measured line fluxes with \texttt{pPXF} using the MILES stellar library, fitting and subtracting the stellar continuum before Gaussian fitting of emission lines. \citet{eftekhari23} adopted values from \citet{gordon23}, who used \texttt{Prospector} with the same MILES library (via FSPS). Fourteen high-S/N hosts employed the \textit{nebular marginalization} mode (similar in concept to \texttt{pPXF}), while eight lower-S/N cases relied on a \texttt{nebemlineinspec} prior based on \textsc{Cloudy} grids. Systematic offsets are thus expected to be minimal, though low-S/N spectra may have larger uncertainties.} Since our primary goal is to investigate the gas-phase metallicity and other physical properties that directly trace star formation in FRB host galaxies, we apply the empirical demarcation proposed by \citet{kauffmann03} to identify and select purely star-forming systems. Galaxies that pass the BPT selection but lie below the star-forming main sequence are still included, as they have measurable emission lines and are therefore not truly quiescent; excluding them would limit our ability to compare FRB hosts with the full background population. Out of 39 galaxies, 10 galaxies lying to the right of this criteria in Figure~\ref{fig:selection} are classified as potential AGN or LINER hosts and are excluded from further analysis. By combining this BPT-based selection with our volume-limited criteria, we construct a final, unbiased sample of star-forming FRB host galaxies suitable for robust metallicity measurements and statistical comparison with the general star-forming galaxy population. This procedure results in a 
remaining sample of $N=29$ hosts out of $39$ (after Step 3 of ``volume-limited'' sample in Table \ref{tab:selection} and  Figure \ref{fig:selection}).

\subsection{Gas-Phase Metallicity Estimate}
\label{ss:metal}

To estimate the gas-phase metallicities (oxygen abundance, $\Oabundance$) of the FRB host galaxies by strong-line indicators \citep[e.g.,][]{maiolino19}, we use  O3N2 index \citep{pettini04},
defined as
\beq
\label{eq:O3N2_definition}
{\rm O3N2} = \log\left(\frac{[\mathrm{OIII}] \lambda5007 / \mathrm{H}\beta}{[\mathrm{NII}] \lambda6584 / \mathrm{H}\alpha}\right).
\eeq
We note that \OIII$\lambda$5007/\Hb~and \NII$\lambda$6584/\Ha~are barely affected by dust because the line pairs are close in wavelength.
Strong-line diagnostics such as O3N2 are calibrated either to direct methods—based on measurements of the electron temperature ($T_e$)—or to photoionization models.
In this work, we adopt an empirical calibration from \citet{curti17} (hereafter C17), who fit relations between various emission line ratios and $T_e$-based metallicity measurements from a sample of HII regions.
Following C17, we compute the oxygen abundance using the O3N2 calibration parameterized as
\beq
\label{eq:O3N2_calibration}
{\rm O3N2} = 0.281 - 4.765 \,x - 2.268 \,x^2,
\eeq
where O3N2 is the diagnostic defined in Eq.~\eqref{eq:O3N2_definition}, and $x \equiv \log({\rm O/H}) -\log({\rm O/H})_\odot$ is the oxygen abundance normalized to the solar value ($12 + \log({\rm O/H})_\odot = 8.69$; \citealt{Zsolar01}). This calibration has been shown to be consistent with direct $T_e$-based methods, with an intrinsic uncertainty of $0.09$ dex in $\log({\rm O/H})$, and is applicable over the range $\Oabundance=7.6$--$8.85$.

We obtained $\Oabundance$ by solving Eq. \eqref{eq:O3N2_calibration} for $x$, and the inferred values, along with other host galaxy information, are summarized in Table~\ref{tab:frb_host_data} in Appendix~\ref{s:fullsample}. We could not estimate $\Oabundance$ for FRB 20190102C due to upper limits on both \NII$\lambda$6584 and \OIII$\lambda$5007, causing this host to be excluded after Step 4 in both ``volume-limited'' and ``no-volume-limit'' samples shown in Table \ref{tab:selection}.

After applying our selection criteria, the final sample consists of $N=27$ hosts (plus one with upper limit; see Step 4 of ``volume-limited'' sample in Table \ref{tab:selection}) with oxygen abundances ranging from $\Oabundance=8.35$--$8.85$, which lies within the valid range of the C17 calibration. The uncertainties in $\log({\rm O/H})$ arising from line flux measurements ($\sim0.01$ dex) are typically much smaller than the intrinsic uncertainty of the C17 calibration ($0.09$~dex), which we therefore adopt as the representative measurement error.
Among hosts that pass our selection, we could only place upper limits on $\Oabundance$ for FRB 20210410D because \OIII$\lambda$5007 was not detected. 
As a result, this host is excluded from the quantitative metallicity comparison with the background galaxy sample discussed in \S \ref{s:result}, although we still include them when comparing $\mstar$ and $\sfr$.

The oxygen abundances for our full FRB host sample (all available hosts without volume-limited selection, $N=31$, plus four upper and lower limits; see the “no-volume-limit” sample in Table \ref{tab:selection}) fall within the valid range of the C17 calibration. Excluding the two hosts with upper or lower limits, the sample shows $\Oabundance = 8.04$–$8.85$ (mean $8.60$, scatter $0.17$) at $z = 0.04$–$0.98$ (mean $0.28$, scatter $0.19$).

\section{Modeling Background Galaxies}
\label{s:model}

To assess the relative importance of stellar mass ($\mstar$), SFR, and gas-phase metallicity in governing FRB production, we require an unbiased and representative sample of star-forming galaxies. Such a sample allows us to test different hypotheses about where FRBs are most likely to occur.
Below, we describe our method for constructing this background comparison galaxy sample.

\subsection{Stellar Mass \& Star Formation Rate}
\label{ss:model_mass_sfr}

To generate a mock galaxy sample of $\mstar$ and $\sfr$ tracing the star-formation main sequence,
we used the public code \texttt{GALFRB}\footnote{\url{https://github.com/loudasnick/GALFRB}} \citep{loudas25}.
This code uses a redshift-dependent galaxy distribution function derived by \citet{leja22} from spectral energy distribution (SED) modeling, trained on the 3D-HST \citep{skelton14} and COSMOS-2015 \citep{laigle16} galaxy surveys covering the redshift range of $z=0.2$--$3$.
The modeling adopts the non-parametric SFHs with \texttt{Prospector} code \citep{leja17,johnson21} to infer stellar masses and SFRs.
The resulting continuous function provides the probability density $\rho_{\rm gal}(\mstar, \sfr \,|\, z)$ of drawing a galaxy with stellar mass $\mstar$ and $\sfr$ at redshift $z$.

Since the underlying training dataset only covers $z=0.2$--$3$, the model is well constrained in this range.
For lower redshifts ($z<0.2$) and very low masses, the code follows the method of \citet{sharma24}, smoothly extending the distribution along the ridge-line of the star-formation main sequence to ensure a realistic extrapolation where direct data are limited.
AGNs are not explicitly included, as is typically the case unless specific AGN templates are applied or sources are flagged in the SED catalog. As a result, the sample is dominated by galaxies lying on or near the star-forming main sequence.
Therefore, we can reasonably assume that the modeled galaxies would satisfy the same BPT-diagram selection criteria applied to the observed sample (see \S \ref{ss:bpt-selection}).
With \texttt{GALFRB}, we generate background galaxies at $0.01 < z < 0.48$ based on the volume-limited selection of the observed sample (\S \ref{ss:vl-selection}), and obtain $\rho_{\rm gal}(\mstar, \sfr)$ following the same fiducial setup as in \citet{loudas25}, with minor modifications (see \S \ref{ss:color}).

\subsection{Gas-Phase Metallicity}
\label{ss:model_metal}

Over the past decades, strong observational evidence has established a correlation between stellar mass and gas-phase metallicity — the so-called mass-metallicity relation (MZR) \citep[e.g.,][]{lequeux79,tremonti04,kewley08}.
Moreover, the MZR has been found to show an additional dependence on $\sfr$, with highly star-forming galaxies tending to be more metal-poor at a given stellar mass. To account for this and reduce the scatter in the MZR, \citet{mannucci10} introduced the fundamental metallicity relation (FMR), which incorporates SFR as an additional parameter:
\beq
\label{eq:mu}
\mu_\alpha = \log  (\mstar) - \alpha \times \log(\sfr)\,,
\eeq
where $\alpha$ is an observationally determined constant between 0 and 1, not a free parameter of our model, chosen to minimize the
scatter in $\mu_\alpha$--$Z$ space \citep{mannucci10,andrews13,curti20}. The FMR appears universally constant up to $z \approx 6$ \citep{sanders21,curti24}. With the $\mstar$ and $\sfr$ values modeled with \texttt{Prospector} through \texttt{GALFRB} (\S \ref{ss:model_mass_sfr}), we assign metallicities to our mock sample using the FMR. Using the FMR is preferable to the MZR alone, as it implicitly accounts for the redshift evolution of the MZR through its dependence on SFR.
Specifically, we adopt the parametrization from \citet{sanders21} (Eq.~10; $\alpha = 0.60$), which is based on $T_e$-based metallicity measurements from an SDSS galaxy sample and is consistent with values reported in the literature ($\alpha \approx 0.55$–$0.7$; \citealt{andrews13,sanders17,curti20}):
\beq
\label{eq:fmr}
\Oabundance = 8.80 + 0.188y - 0.220y^2 - 0.0531y^3 ,
\eeq
where $y=\mu_{0.60}(\mstar,\sfr)-10$. To account for the intrinsic scatter observed in real galaxies, we added an intrinsic scatter of $0.05$ dex to the $\Oabundance$; this value reflects the scatter in the observed FMR and is independent of the uncertainties in the mock sample.
We then generated $\Oabundance$ values for the mock galaxies using this relation and scatter.

Because the $\mstar$ and $\sfr$ values in the FMR (Eq.~\ref{eq:fmr}) are calibrated on SDSS galaxies modeled with parametric SFHs (pSFHs), while our \texttt{GALFRB} catalog relies on estimates based on non-parametric SFHs (npSFHs) with \texttt{Prospector}, we correct for the systematic offset between these approaches as follows. 

As discussed by \citet{leja22}, galaxies modeled with npSFHs using \texttt{Prospector} tend to yield stellar masses systematically higher by $\sim$0.2~dex compared to those derived from pSFH models, an effect robustly confirmed for galaxies at $0.7 < z < 1.3$ (see Figure 9 in Section 6.2 of \citealt{leja22} and Figure \ref{fig:fig9leja22} in Appendix \ref{ss:leja22}). The offset arises primarily from the difference in SFH parameterization: non-parametric models allow for older stellar populations and more flexible recent SFHs, resulting in systematically higher inferred mass-to-light ratios. Consequently, the effect is unlikely to have a strong redshift dependence, at least around $z \lesssim 1$.
We therefore apply a uniform $0.2$~dex correction to the stellar masses of our model galaxies, independent of redshift, when applying the FMR calibrated with pSFHs,
$\log(M_{\star,\rm pSFH}) = \log(M_{\star,\rm npSFH}) - 0.2$~dex. 

Regarding SFR, because a direct npSFH–pSFH SFR comparison is still poorly constrained, any conversion between the two remains inherently uncertain. The closest available reference is \citet{leja22}, who compare npSFH SFRs to UV+IR–based SFRs—an approximate proxy for pSFH estimates—and report offsets of $-0.1$ to $-1$ dex for galaxies, with an average shift of $\sim -0.3$ dex at $0.7 < z < 1.3$ (Figure~9 of \citealt{leja22}; see also Figure~\ref{fig:fig9leja22} in Appendix~\ref{ss:leja22}). Given both this uncertainty and the lower redshift of our FRB host sample ($\langle z \rangle \approx 0.3$), we do not apply an SFR correction in our main analysis (but see \S\ref{sss:FMR_correction} for a test using a uniform SFR shift). Instead, we conservatively adopt the stellar-mass correction alone as our fiducial approach.

These considerations ensure that our metallicity modeling remains consistent with the empirical FMR calibration with pSFHs while preserving the internally consistent \texttt{Prospector}-based ($\mstar$, $\sfr$) npSFH estimates within the mock galaxy sample. All quantities in our analysis (both for FRB host galaxy data and the model) are, unless otherwise noted, calibrated to the npSFH scale
derived with \texttt{Prospector}.

\begin{figure*}
\centering
\includegraphics[width=17cm]{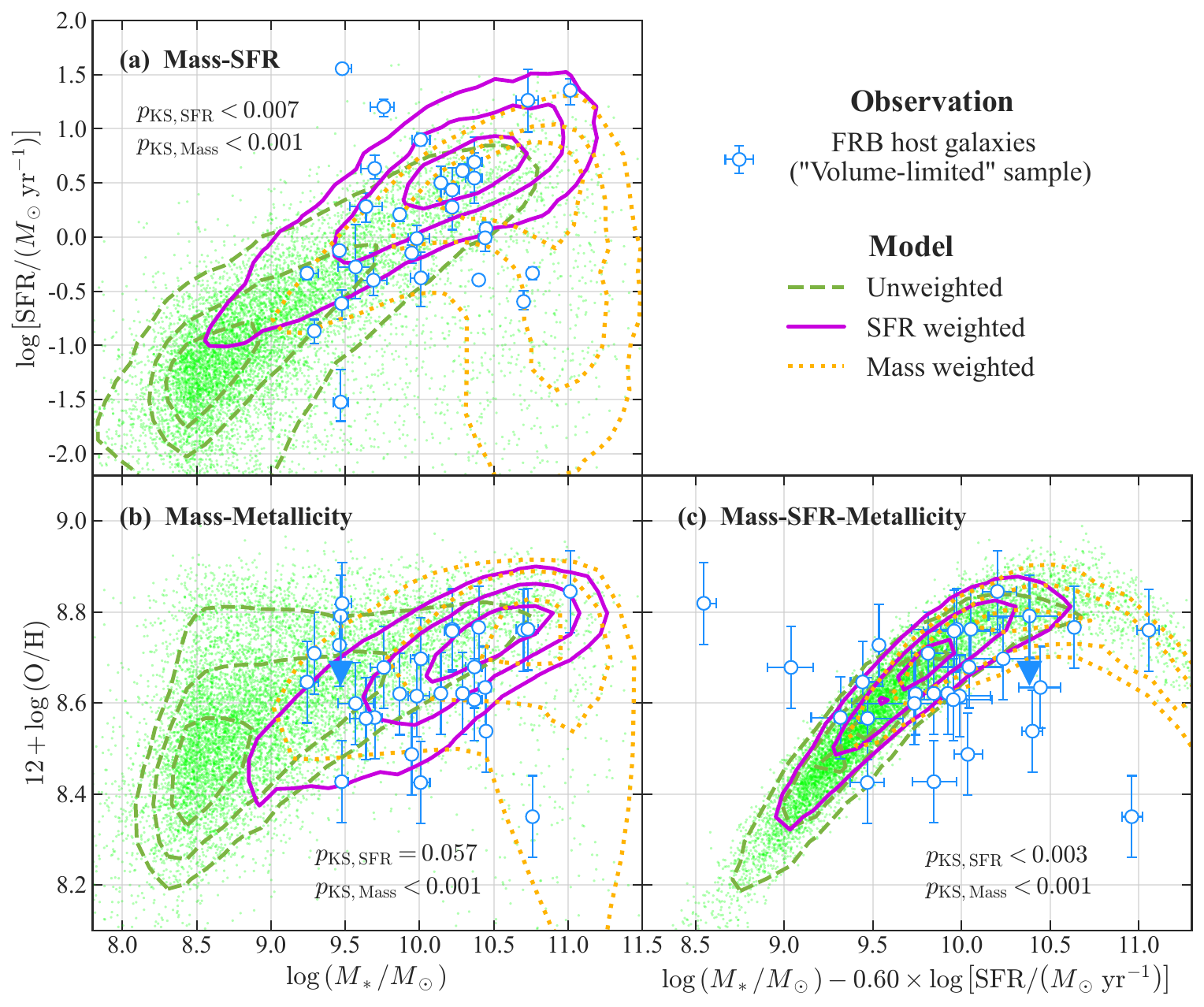}
\caption{Comparison of FRB host galaxy properties with the general galaxy population.
(a) Stellar mass vs. SFR ($N=29$; see Step 3 of ``volume-limited'' sample in Table \ref{tab:selection}),
(b) Stellar mass vs. gas-phase metallicity ($N=27$, plus one upper limit; see Step 4 of ``volume-limited'' sample in Table \ref{tab:selection}),
(c) Fundamental metallicity relation ($N=27$, plus one upper limit; same as panel b).
FRB host galaxies are shown as open blue circles. Light green points represent the control sample. Green dashed, purple solid, and orange dotted contours show the original, SFR-weighted, and mass-weighted control sample distributions, respectively. Contours correspond to the 20\%, 50\%, and 80\% cumulative density levels. The 2D-KS $p$-values are indicated in each panel.}
\label{fig:dist_2d}
\end{figure*}

\begin{figure*}
\centering
\includegraphics[width=17cm]{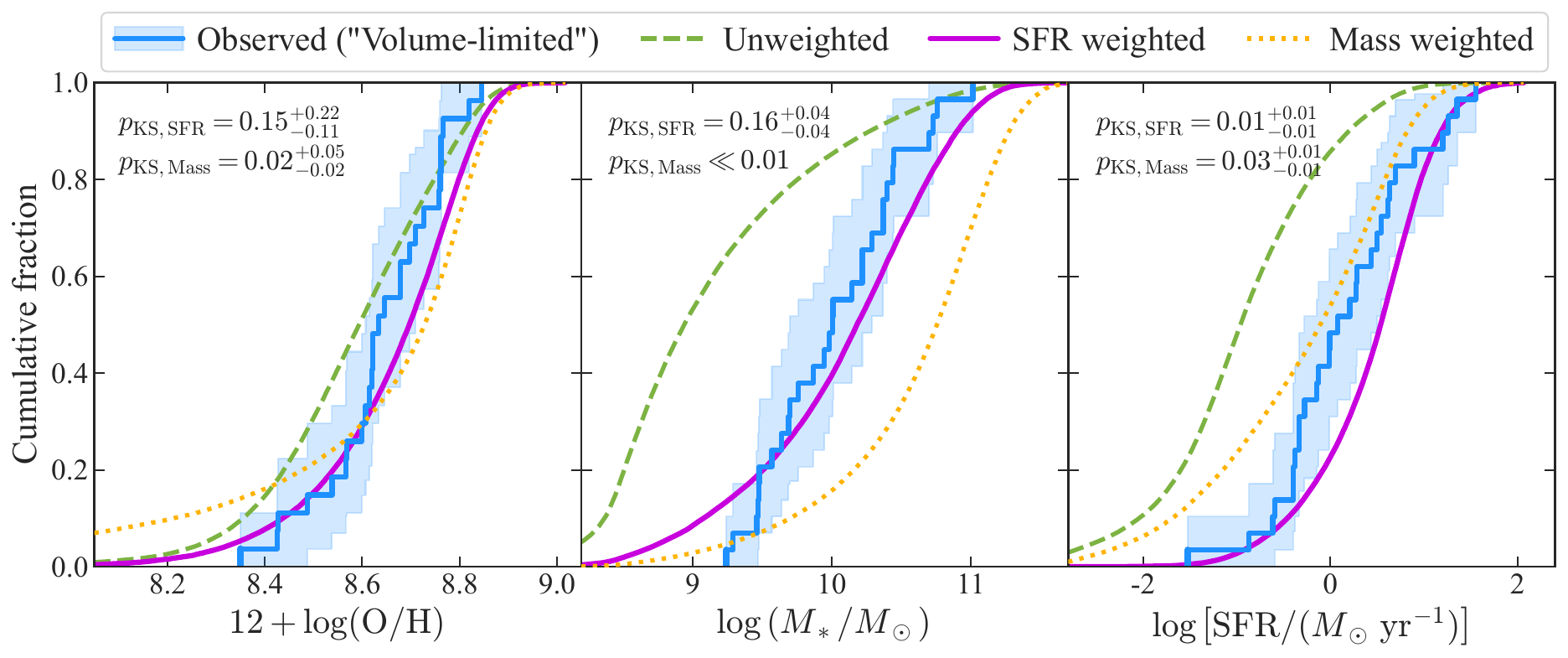}
\caption{Comparison of cumulative distributions of FRB host galaxy properties (“volume-limited” sample defined in Table \ref{tab:selection} and used in Figure \ref{fig:dist_2d}): from left to right, gas-phase metallicity ($N=27$), stellar mass ($N=29$), and SFR ($N=29$). Solid blue histograms show the CDFs of the FRB hosts, and the shaded blue areas indicate the 95\% confidence intervals estimated by bootstrapping. Colored curves show the different models discussed in the text: the unweighted model (dashed green), the SFR-weighted model (solid purple), and the mass-weighted model (dotted orange). The 1D-KS $p$-values are indicated in each panel.}
\label{fig:dist_1d}
\end{figure*}

\subsection{Color}
\label{ss:color}

\citet{loudas25} use a model that links the SFR of each galaxy to its rest-frame mass-to-light ratio ($\mstar/L_r$) through the rest-frame optical color $(g-r)_\mathrm{rest}$. They draw on the SDSS+WISE MAGPHYS catalog \citep{chang15}, which contains SED fits to nine-band photometry for roughly $10^6$ galaxies at $z<0.2$, to extract the conditional probability density $P((g-r)_\mathrm{rest} \,|\, \sfr)$ and then sample colors for all mock galaxies.

However, while this provides a useful framework, the model neglects the fact that galaxy color is not determined solely by SFR—it also strongly correlates with stellar mass. As an improvement over the \citet{loudas25} model, we extract the joint conditional probability density $P((g-r)_\mathrm{rest} \,|\, \mstar, \sfr)$ instead. This modification helps to retain faint, low-mass but high-SFR galaxies that might otherwise be excluded by an apparent magnitude cut. Since the background galaxy distribution is typically weighted by SFR or stellar mass, this correction does not significantly alter the overall shape of the comparison sample. Nonetheless, we implement this adjustment to better reflect the underlying physical correlations in the model.
Additionally, since the $\mstar$ and $\sfr$ values in the SDSS+WISE MAGPHYS catalog are derived from SED-fitting using  pSFHs, while our \texttt{GALFRB} catalog relies on those derived from npSFHs with \texttt{Prospector}, we correct for the systematic offset in $\mstar$ measured between these approaches using the same method described in \S \ref{ss:model_metal} when assigning colors to the modeled galaxies. 

After generating the sample as described in \S \ref{ss:model_mass_sfr}, we apply an absolute magnitude cut of M$_r < -17.5$, based on the volume-limited selection (\S \ref{ss:vl-selection}), to obtain the final background galaxy sample for comparison with the observed data.

\section{Result}
\label{s:result}

We now have the observed FRB host galaxy sample, with measurements of $\mstar$ and $\sfr$, along with $\Oabundance$  independently derived from emission line measurements, as described in \S \ref{s:data}. In parallel, we construct a background (control) galaxy sample selected using the same criteria (\S \ref{ss:vl-selection} and \S \ref{ss:bpt-selection}), as detailed in \S \ref{s:model}. This setup enables us to investigate whether FRB hosts are representative of the general star-forming galaxy population or show systematic differences.

To explore this, we examine three key 2D planes defined by combinations of $M_\star$, SFR, and gas-phase metallicity:
(a) $\mstar$–SFR, (b) $\mstar$–metallicity, and (c) $\mu_{0.60}(\mstar,\sfr)$–metallicity (see Eq.~\ref{eq:mu} for the definition of $\mu_{0.60}$).
Figure~\ref{fig:dist_2d} shows these distributions: FRB hosts are marked as blue circles (with filled symbols denoting metallicity upper or lower limits), while the control sample appears as light green points. Green dashed contours indicate the (unweighted) distribution of the control sample.

To quantify these differences, we perform a two-dimensional (2D) Kolmogorov–Smirnov (KS) test in each of the three planes. Specifically, we compare the FRB host distribution $\rho_{\rm FRB}$ with the control sample distribution $\rho_{\rm gal}$ under different weighting schemes: unweighted, SFR-weighted ($\sfr \times
\rho_{\rm gal}$), and stellar-mass–weighted ($\mstar \times\rho_{\rm gal}$). This allows us to evaluate the likelihood that the observed FRB hosts could be randomly drawn from each of these background distributions. Throughout this work, we adopt a KS test p-value threshold of $p_{\rm KS} = 0.05$ as the statistical significance.

As shown in Figure~\ref{fig:dist_2d}, the FRB host galaxies are overall broadly consistent with the SFR-weighted background population (purple solid contour).
In the $\mstar$–$Z$ plane shown in Panel (b), the hosts follow the MZR seen in the control sample, with a reasonably high 2D-KS p-value ($p_{\rm KS, SFR}\sim0.06$). This suggests that, in terms of metallicity at a given stellar mass, FRB hosts do not show strong systematic offsets.
In contrast, in the $\mstar$–SFR plane shown in Panel (a), some FRB hosts fall slightly below the star-forming main sequence traced by the SFR-weighted background galaxies, reflected in a low 2D-KS p-value ($p_{\rm KS, SFR}<0.007$). Similarly, in the $\mu_{0.60}(\mstar,\sfr)$–metallicity plane shown in Panel (c), FRB hosts tend to have deviations in metallicities at given $\mu_{0.60}(\mstar,\sfr)$ than predicted by the SFR-weighted background, leading to a low KS p-value ($p_{\rm KS, SFR}<0.003$).

\begin{figure*}
\centering
\includegraphics[width=17cm]{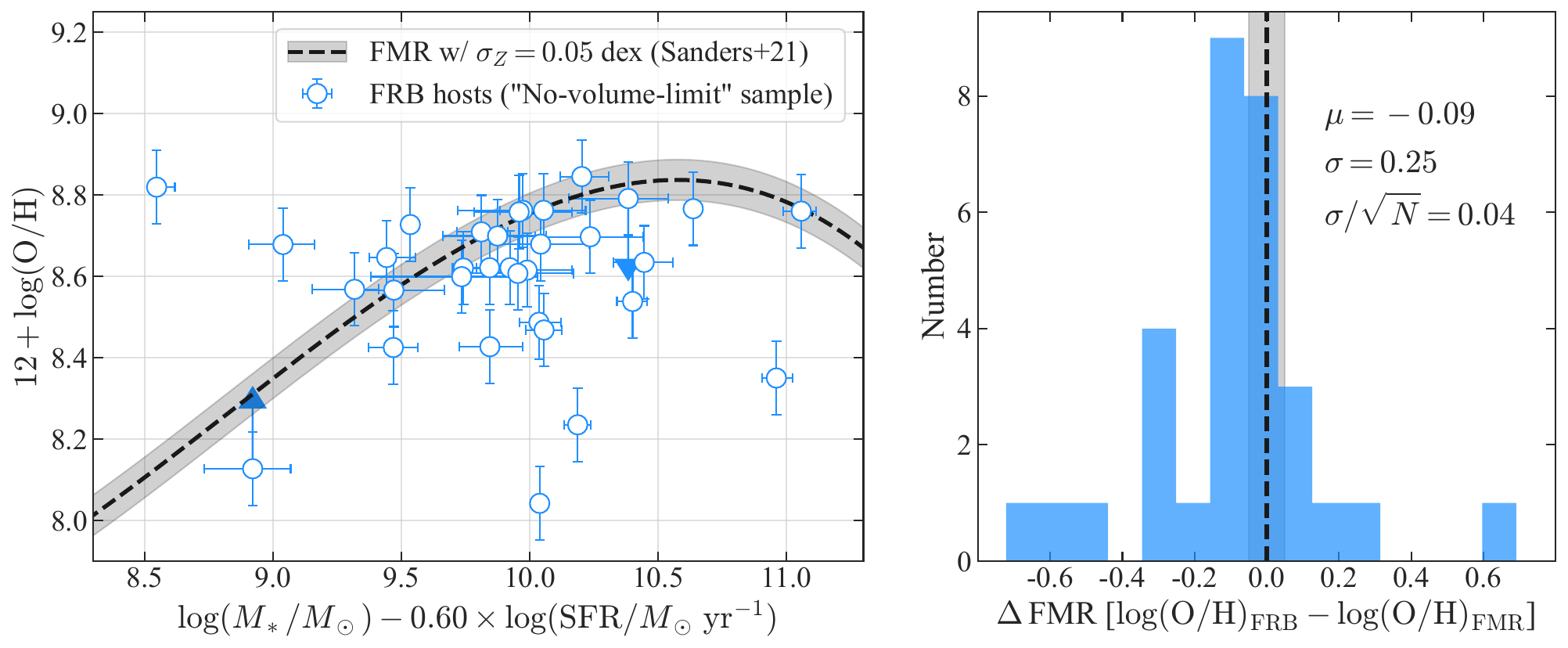}
\caption{{\it Left}: FRB host galaxies (
$N=31$, plus 2 upper/lower limits) selected based on the BPT diagram (see ``no-volume-limit'' sample in Table \ref{tab:selection} and \ref{tab:frb_host_data} in Appendix \ref{s:fullsample} 
for the details), compared with the fundamental metallicity relation (FMR) from \citet{sanders21}, which has been robustly confirmed for galaxies at $z=0$–3. The FMR is shown as a dashed black line with a dispersion of $0.05$ dex along the metallicity axis. {\it Right}: Histogram of metallicity offsets from the FMR for the FRB host sample, using only the $N=31$ hosts with metallicity measurements. The mean ($\mu$), standard deviation ($\sigma$) and standard error ($\sigma/\sqrt{N}$) are indicated in the panel.
}
\label{fig:fmr}
\end{figure*}

Figure~\ref{fig:dist_1d} presents the one-dimensional (1D) cumulative distributions of $\Oabundance$, $\mstar$, and SFR.
Quantitatively, the SFR-weighted model shows good agreement with the FRB host distributions for metallicity and stellar mass, with significantly high 1D-KS p-values:
for metallicity, $p_{\rm KS, SFR} = 0.15$, and for stellar mass, $p_{\rm KS, SFR} = 0.16$.
This indicates that FRBs do not show a significant preference for lower or higher metallicity environments compared to typical star-forming galaxies, which is a notable result, as this is the first time the metallicity distribution of FRB hosts is found to be consistent with an SFR-weighted control sample with proper volume-limited selection.
However, the SFR distribution still shows a mild but significant deviation (1D-KS $p_{\rm KS,SFR}=0.01$), indicating that FRB hosts have slightly lower SFRs than expected under pure SFR weighting.
This could potentially contribute to the deviations observed from the FMR (see \S \ref{ss:sfrdriven} for discussion).

As shown in panel (c) of Figure~\ref{fig:dist_2d}, FRB hosts systematically deviate from the FMR.
Here, we examine these deviations in more detail and discuss their possible implications.
Given the redshift invariance of the FMR, we relax the volume-limited sample selection and instead include all observed FRB host galaxies that satisfy the BPT-diagram selection (see ``no-volume-limit'' sample in Table \ref{tab:selection}), to maximize the sample size.
The left panel of Figure~\ref{fig:fmr} shows the observed FRB hosts compared to the FMR used to assign metallicities in our simulations (Eq.~\ref{eq:fmr}).
For each host, we compute its offset from the FMR at the corresponding $\mu_{0.60}$ value, and the distribution of these offsets is shown in the right panel of Figure~\ref{fig:fmr}.
Considering only detections ($N=31$), we find that the majority (72\%) of FRB hosts lie below the FMR, with an average metallicity offset of $-0.09$~dex and a scatter of $0.24$~dex. The standard error of the mean is $0.04$~dex, indicating weak evidence ($\sim2\sigma$) that FRB hosts are systematically lying below the FMR. 

When restricting the analysis to the ``volume-limited'' sample in Table \ref{tab:selection} (i.e., galaxies selected by both redshift and BPT criteria; see Figures \ref{fig:dist_2d} and \ref{fig:dist_1d}), the offset decreases to  $-0.04\pm0.04$~dex and becomes less significant (1$\sigma$). However, since the FMR is intrinsically redshift-independent, this smaller offset likely reflects the reduced sample size ($N=27$) rather than a physical difference. We therefore adopt the ``no-volume-limit'' sample as our fiducial case for FMR comparison (see also \S \ref{ss:offset}), as it provides a more statistically robust result.

In summary, FRB hosts are largely consistent with an SFR-weighted background of star-forming galaxies and the MZR. Yet, there could be subtle but systematic deviations from the SFR main sequence and the FMR, visible both in 2D and in the 1D cumulative distributions.

\section{Implications}
\label{s:discussion}

\subsection{Do FRB Progenitors Depend on Metallicity}

Our analysis shows that FRB host galaxies broadly follow the MZR of star-forming galaxies when properly weighted by SFR (\S \ref{s:result}). This is consistent with the trend reported by \citet{heintz20} based on seven hosts, although their sample size was limited and no volume-limited selection was applied to either the observed or comparison galaxies. Here, we extend the analysis to the largest to-date ($N=27$) volume-limited sample with a consistently selected control population.

Importantly, we do not find evidence for a sharp low-metallicity cutoff within our primary volume-limited sample ($0.01 < z < 0.48$, $M_r < -17.5$), suggesting that FRBs occur over a broad metallicity range. To test whether such a cutoff might exist below the luminosity threshold of this selection, we constructed an alternative, deeper volume-limited sample with fainter control galaxies ($0.01 < z < 0.17$, $M_r < -15.1$). This supplementary sample, not included in Table~\ref{tab:selection}, contains fewer hosts ($N=10$ with metallicity measurements after Steps~0--4), but the comparison yielded a similar result: FRB hosts still follow the SFR-weighted MZR of the control sample with a KS test p-value of $p_{\mathrm{KS,SFR}} \approx 0.05$, indicating a marginally significant consistency between the two distributions. 
This consistency demonstrates the robustness of our conclusion, likely due to the minimal contribution of low-luminosity (low-SFR) galaxies in the SFR-weighted comparison.

Additionally, one of our sample FRB 20221029A outside the volume limit (``no-volume-limit'' sample in Table \ref{tab:selection}) shows a remarkably low metallicity of $\Oabundance=8.04_{-0.03}^{+0.04}$, which is comparable to the lowest metallicities known for FRBs in dwarf hosts reported in the literature, such as FRB 20121102A \citep[$\Oabundance=8.0\pm0.1$;][]{bassa17} and FRB 20230708A \citep[$\Oabundance\approx 7.99$–8.3;][]{august25}. Together, these results indicate that FRBs could occur across a wide range of host metallicities, and that metallicity is unlikely to be a primary regulator of progenitor formation.

This stands in contrast to the interpretation by \citet{sharma24} that proposed a possible metallicity dependence in FRBs, inferred indirectly from an apparent lack of low-mass hosts in their flux-limited samples. Their argument, however, was based only on the stellar mass distribution, combined with simplified assumptions about optical detectability, without directly analyzing metallicity. Subsequent studies that incorporated both stellar mass and SFR with more realistic treatment of detection thresholds  \citep{loudas25,horowicz25} also found no strong evidence for a strict lower stellar mass boundary—but likewise did not examine metallicity directly.

Nevertheless, we recognize that metallicity could still shape FRB progenitor pathways. For example, higher metallicities can drive stronger stellar winds, favoring neutron star remnants over black holes in single-star evolution, or increase the frequency of stellar mergers in binaries, potentially forming magnetars through magnetic blue straggler channels \citep{sharma24}. Although our results suggest that metallicity does not impose a strict cutoff fully suppressing FRB production in low-metallicity systems, given the uncertainties and lack of quantitative models, we cannot rule out that it modulates the relative contribution or efficiency of different formation channels (e.g., single versus binary pathways).

Encouragingly, while metallicity may influence these relative contributions of different progenitor channels and add complexity, we find that FRBs remain broadly linked to star formation itself. If FRBs indeed trace star formation without a strong preference for high metallicity, they should also form and be detectable even in low-metallicity, high-redshift galaxies in the early Universe \citep{zhang18,hashimoto20}. This preserves their potential as unique probes of helium reionization at $z\sim3$–4 \citep{caleb19,linder20}, cosmic hydrogen reionization at $z\gtrsim6$ \citep{yoshiura18,hashimoto21,heimersheim22}, and the cosmic baryon distribution \citep{lee22,connor25}—especially in the upcoming era of the Square Kilometre Array (SKA; \citealt{dewdney09}).

\subsection{Offset from the FMR}
\label{ss:offset}
While we find FRB hosts consistent with the MZR, at the same time, we find marginal $\sim2\sigma$ evidence—an average offset of $-0.09\pm0.04$~dex—that FRB hosts lie systematically below the FMR, which connects stellar mass, SFR, and metallicity (\S \ref{s:result}).
In what follows, we first examine whether this offset can be accounted for by systematic corrections to both stellar mass and SFR from npSFHs (\S~\ref{sss:FMR_correction}) or by adopting a different FMR (\S~\ref{sss:FMR_alpha_varied}). We then explore the possibility that it reflects intrinsically low SFRs in FRB hosts and their potential physical implications (\S~\ref{sss:FMR_implication}).

\label{ss:sfrdriven}
\subsubsection{Impact of Mass and SFR Systematic Calibrations}
\label{sss:FMR_correction}

In the main analysis, we applied a uniform $0.2$~dex correction to stellar masses to account for systematic differences between values derived from npSFHs with Prospector and those obtained using pSFH methods (see \S \ref{ss:model_metal}). To test the potential impact of also correcting SFRs on the FMR offsets, we have performed an analysis accounting for both mass and SFR shifts based on the galaxy’s location in the mass–SFR plane based on \citet{leja22}. The details of this procedure, including the adopted shift vectors and comparison with the FMR on the pSFH scale, are provided in Appendix~\ref{ss:leja22}. We find that applying these corrections slightly reduces the metallicity offset of FRB hosts relative to the FMR from $-0.09 \pm 0.04$ dex to $-0.06 \pm 0.04$ dex (1.5$\sigma$), but the hosts remain systematically below the relation (see Figure \ref{fig:fmr_msfr_shifted} in Appendix \ref{ss:leja22}). This indicates that the observed offset cannot be fully explained by potential mass–SFR calibration and may indeed exist.

To further explore the impact of systematic SFR differences, we consider the uniform SFR shift that would be required to eliminate the FMR offset. We estimate this shift to be $\sim -0.5$ dex, which lies within the broad range of reported npSFH–pSFH SFR offsets ($-0.1$ to $-1$ dex), though it is relatively large compared to the typical shift of $\sim -0.3$ dex for galaxies. Therefore, the observed metallicity offset cannot be explained by the uniform mass correction alone and would require a comparatively large uniform SFR adjustment to be removed.

\subsubsection{Impact  of Using a Different FMR}
\label{sss:FMR_alpha_varied}

In the main analysis, we compared the FMR with $\alpha = 0.6$ derived by \citet{sanders21} (see Section~\ref{ss:model_metal}). To test the potential impact of adopting alternative values of $\alpha$ from the literature, we performed an additional analysis using $\alpha = 0.55$ from \citet{curti20} (``Global sample'' in their Table~A1). The details of this procedure are provided in Appendix~\ref{ss:curti20}. We find that applying these corrections reduces the metallicity offset of FRB hosts relative to the FMR from $-0.09 \pm 0.04$~dex to $-0.03 \pm 0.04$~dex (0.8$\sigma$), suggesting that the choice of FMR may help reconcile the apparent offset (see Figure \ref{fig:fmr_alpha055} in Appendix \ref{ss:curti20}). However, when both stellar-mass and SFR corrections are additionally applied (as in Section~\ref{sss:FMR_correction}), the offset becomes $-0.07 \pm 0.04$~dex (1.8$\sigma$), indicating that a residual offset may still persist.

\subsubsection{Offset Driven by Intrinsically Low SFR?}
\label{sss:FMR_implication}
Although the true statistical significance may be influenced by the model uncertainties discussed above (\S~\ref{sss:FMR_correction} and \ref{sss:FMR_alpha_varied}), treating the offset as real allows us to explore possible physical causes. The shift is unlikely to be driven by unusually low metallicities: if low metallicity were responsible, we would expect a similar systematic offset in the mass–metallicity plane, which is not observed. The remaining possibility is an offset along the $\log M_\star-0.6 \log \sfr$ axis (i.e., in mass and/or SFR). Indeed, 1D KS test (Figure \ref{fig:dist_1d}) shows that FRB hosts have significantly lower SFRs than SFR-weighted comparison galaxies of similar stellar mass, while their metallicities and stellar masses remain consistent. This suggests that the primary driver of the FMR offset, if genuine, is suppressed star formation activity at fixed mass and metallicity rather than metallicity deficiency.

The FMR is expressed as a function of $\log M_\star - 0.6 \log \mathrm{SFR}$ \citep[e.g.,][]{mannucci10,sanders21}. Galaxies with SFR lower by $\sim0.5$~dex shift their FMR coordinate rightward by $\sim0.3$~dex, producing an apparent metallicity deficit of $\sim0.1$~dex relative to the FMR curve, consistent with the observed mean offset of FRB hosts ($\sim 0.09$~dex).

A plausible explanation for such SFR deficiency is that some FRB hosts are caught in a post-starburst or quenching phase, shortly after a merger-induced starburst has faded. In this stage, SFR drops rapidly while stellar mass and metallicity remain high. Local post-merger galaxies are also known to lie $\sim0.1$–$0.2$~dex below the FMR \citep[e.g.,][]{gronnow15,bustamante20}, consistent with our observed offset, although their SFR evolution is less well constrained observationally. This scenario naturally accounts for both the low SFRs and the retention of high stellar mass and metallicity. Furthermore, in this scenario, we might expect to see kinematically disturbed signatures of recent interactions, as reported in some FRB host galaxies through \HI~studies \citep{Michalowski21, hsu23,glowacki23}, and in an FRB hosted by a compact galaxy group \citep{gordon24}.

Starburst activity triggered by galaxy mergers typically lasts and is quenched over a timescale of $\sim$100–500 Myr \citep[e.g.,][]{dimatteo07,ellison24}.
The observed delay-time distribution of FRBs—estimated as $t^{-1.9 \pm 1.1}$ with a minimum delay of $\sim110^{+250}_{-20}$~Myr (\citealt{law24}, see also \citealt{gordon23})—is broadly compatible with this timescale. If so, FRB progenitors could form $\gtrsim100$~Myr after the burst episode, and the currently low SFRs of their host galaxies may reflect the rapid post-starburst decline in star formation, consistent with the required evolutionary timescale. In this scenario, FRB progenitors are created in the interaction-driven starburst phase, and a subset of them eventually produce FRBs. Qualitatively, the non-negligible minimum delay time ($\sim100$--$500$~Myr) implied by this scenario favors progenitor channels involving binary evolution \citep[e.g., ][]{Zhang2025} rather than prompt core-collapse supernovae, although CCSNe could still contribute in systems with shorter delays. Additional host galaxy observations are needed to confirm this interpretation.

\section{Summary \& Conclusions}
\label{s:conclusion}

We present the largest uniform study of FRB host galaxies to date, comprising 31 hosts in total (out of a parent sample of 45), including a volume-limited subsample of 28, with consistent, independent measurements of gas-phase oxygen abundances using strong emission-line O3N2 diagnostics with C17 calibration. This enables a robust comparison to galaxy scaling relations and control samples. Our main conclusions are as follows:

\begin{itemize}
\item 
FRB hosts broadly follow the star-formation-weighted mass–metallicity relation, suggesting that metallicity itself does not strictly limit FRB progenitor formation. We detect no sharp metallicity threshold within our volume-limited sample, with observed metallicities spanning $\Oabundance \approx 8.3$--$8.8$. Even beyond the volume limit, some FRBs occur in very low-metallicity massive hosts (e.g., FRB 20221029A with $\Oabundance = 8.04$), comparable to the lowest metallicities found in dwarf hosts \citep{bassa17,august25}. These findings imply that FRBs can form in typical star-forming environments across a wide metallicity range, contrasting with the possible metallicity dependence inferred indirectly from host mass distributions by \citet{sharma24}

\item 
The absence of a strict metallicity dependence suggests that metallicity does not impose a hard threshold for FRB progenitor formation, though it may still influence the efficiency or relative contribution of different formation channels. 
While this apparent absence of metallicity dependence may
seem to complicate progenitor modeling, it is encouraging that this also implies that FRBs can arise even in the low-metallicity environments characteristic of early galaxies, supporting their potential as powerful probes of the high-redshift Universe and the cosmic baryon distribution in the near future.

\item 
We also find marginal  ($\sim$2$\sigma$) evidence that FRB hosts are offset by $-0.09 \pm 0.04$ dex below the fundamental metallicity relation (FMR) with our fiducial model. After accounting for potential model uncertainties, including effects of mass and SFR calibrations and the choice of FMR, the offset is reduced to $-0.06 \pm 0.04$~dex (1.5$\sigma$) and $-0.03 \pm 0.04$~dex (0.8$\sigma$), respectively, which may partially/fully explain the observed offsets. Nevertheless, given the remaining model uncertainties, the offset might still exist.

\item The offset from FMR, if real, is likely driven by slightly lower SFRs at fixed mass and metallicity, rather than an intrinsic metal deficiency. Such an offset is also observed in local post-merger galaxies \citep{gronnow15,bustamante20}. A plausible interpretation is that some FRBs occur in post-starburst or quenching systems, where a starburst triggered by interactions has faded, but stellar mass and metallicity remain high. In this scenario, progenitors form during the burst and produce FRBs with a delay of $\sim$100–500 Myr \citep{dimatteo07,ellison24}. Such a delay is broadly consistent with observed delay-
time distributions and favoring delayed channels such as binary evolution over core-collapse supernovae \citep{law24}.  We predict that FRBs may preferentially occur in host galaxies which exhibit kinematic signatures of recent interactions, such as tidal features, disturbed velocity fields, or double nuclei. Additional host galaxy observations are needed to test this qualitative interpretation and prediction.

\end{itemize}

Future work could extend our analysis in several directions. (i) Increasing the sample size and constructing a volume-limited host galaxy catalog will be critical for robust statistical comparisons with control populations. (ii) Consistent use of emission-line diagnostics across all samples is essential to minimize systematic uncertainties in metallicity measurements. (iii) Furthermore, although not addressed in this work, potential biases introduced by radio selection in FRB detection could be carefully assessed. (iv) Some well-localized FRBs, relative to their host sizes, are found in the outer regions of their hosts \citep[e.g.,][]{ravi19,kirsten22,gordon23,sharma24,woodland24,shah25,gordon25}. Our analysis is limited to global (galaxy-averaged) properties, and future work combining precise localizations with spatially resolved spectroscopy or IFU observations to measure metallicity {\it at the precise FRB sites} will be key to testing whether progenitors favor locally metal-poor or metal-rich environments. (v) Finally, developing theoretical models that predict or explore metallicity-dependent FRB formation channels will be crucial for interpreting future observational results.

\begin{acknowledgments}
We thank the referee, Wen-Fai Fong, for her thorough review and numerous insightful comments, which have helped to clarify and strengthen the paper. We thank Alexa Gordon and Kritti Sharma for generously sharing detailed information on their SED fitting analyses. We are grateful to Joel Leja for providing the numerical data on the calibration between \texttt{Prospector} and traditional SED modeling methods.  SY thanks Kazumi Kashiyama and Tomonori Totani for valuable comments on our early results. SY also thanks Yu-An Chen and Seong-Jin Kim for discussions regarding galaxy color modeling. SY acknowledges the support from the National Science and Technology Council of Taiwan (NSTC) through grant numbers 113-2112-M-005-007-MY3 and 113-2811-M-005-006-. TH acknowledges the support from the NSTC through grants 113-2112-M-005-009-MY3, 113-2123-M-001-008-, 111-2112-M-005-018-MY3, and the Ministry of Education of Taiwan through a grant 113RD109. HK acknowledges the support from KAKENHI (23KJ2148, 25K17444) through the Japan Society for the Promotion of Science. TG acknowledges the support of the NSTC through grants 108-2628-M007-004-MY3, 111-2112-M-007-021, and 112-2123-M-001-004-.
\end{acknowledgments}

\vspace{5mm}

\appendix

\section{Host Galaxy Properties}
\label{s:fullsample}

We summarize the FRB host galaxy properties used in this study in Table~\ref{tab:frb_host_data}, including redshift, stellar mass, SFR, and oxygen abundance derived from the O3N2 calibration ($12 + \log(\mathrm{O/H})$). We include the parent sample ($N=45$) that can be plotted on the BPT diagram (i.e., those passing Selection Step~1 in Table~\ref{tab:selection}).
We also indicate whether each host passes our sample selections for ``volume-limited'' and ``no-volume-limit'' samples described in \S~\ref{s:data}. Errors on $\Oabundance$ represent 1$\sigma$ lower and upper uncertainties propagated from the measurement errors of the emission line intensities; when plotting, these are replaced by the intrinsic scatter of the C17 calibration ($0.09$ dex) if smaller.

\begin{longtable*}{ccccccc}
\caption{FRB host galaxy properties used in this study} 
\label{tab:frb_host_data} \\
\hline\hline
FRB Name &  $z$ & $\log\mstar$ & $\log\sfr$ &
$\Oabundance$ & VL-Selected? & BPT-Selected?\\
 &  & ($M_\odot$) & ($M_\odot\,$yr$^{-1}$) & 
 (\S \ref{ss:metal}) & (\S \ref{ss:vl-selection}) & (\S \ref{ss:bpt-selection})\\
\hline
\endfirsthead

\multicolumn{7}{c}%
{\tablename\ \thetable\ -- \textit{continued from previous column}} \\
\hline\hline
FRB Name &  $z$ & $\log\mstar$ & $\log\sfr$ &
$\Oabundance$ & VL-Selected? & BPT-Selected? \\
 &  & ($M_\odot$) & ($M_\odot\,$yr$^{-1}$) & 
 (\S \ref{ss:metal}) & (\S \ref{ss:vl-selection}) & (\S \ref{ss:bpt-selection}) \\
\hline
\endhead

\hline 
\endfoot

\hline
\endlastfoot

\hline
20220319D & 0.011 $^{a}$ & $10.101^{+0.002}_{-0.004}$ $^{a}$ & $-0.373^{+0.022}_{-0.022}$ $^{a}$ & 
$8.717^{+0.003}_{-0.003}$ $^{b}$ & Y& N\\
20231120A & 0.037 $^{a}$ & $10.399^{+0.002}_{-0.002}$ $^{a}$ & $-0.396^{+0.020}_{-0.028}$ $^{a}$ & 
$8.766^{+0.005}_{-0.005}$ $^{b}$ & Y& Y\\
20220207C & 0.043 $^{a}$ & $9.948^{+0.031}_{-0.033}$ $^{a}$ & $-0.147^{+0.070}_{-0.091}$ $^{a}$ &
$8.487^{+0.008}_{-0.008}$ $^{b}$ & Y& Y\\
20220509G & 0.089 $^{a}$ & $10.701^{+0.013}_{-0.009}$ $^{a}$ & $-0.595^{+0.100}_{-0.074}$ $^{a}$ & 
$8.760^{+0.004}_{-0.004}$ $^{b}$ & Y& Y\\
20230124A & 0.094 $^{a}$ & $9.461^{+0.004}_{-0.003}$ $^{a}$ & $-0.124^{+0.023}_{-0.023}$ $^{a}$ &
$8.727^{+0.021}_{-0.020}$ $^{b}$ & Y& Y\\
20220914A & 0.114 $^{a}$ & $9.243^{+0.078}_{-0.041}$ $^{a}$ & $-0.334^{+0.046}_{-0.058}$ $^{a}$ & 
$8.646^{+0.019}_{-0.020}$ $^{b}$ & Y& Y\\
20230628A & 0.127 $^{a}$ &$9.292^{+0.032}_{-0.030}$ $^{a}$ & $-0.867^{+0.106}_{-0.118}$ $^{a}$ & 
$8.709^{+0.0006}_{-0.0006}$ $^{b}$ & Y& Y\\
20220920A & 0.158 $^{a}$ & $9.867^{+0.013}_{-0.011}$ $^{a}$ & $0.209^{+0.051}_{-0.062}$ $^{a}$ & 
$8.620^{+0.008}_{-0.008}$ $^{b}$ & Y& Y\\
20221101B & 0.239 $^{a}$ & $11.207^{+0.027}_{-0.018}$ $^{a}$ & $1.089^{+0.141}_{-0.188}$ $^{a}$ & 
$8.649^{+0.127}_{-0.150}$ $^{b}$ & Y& N\\
20220825A & 0.241 $^{a}$ & $10.008^{+0.062}_{-0.057}$ $^{a}$ & $0.898^{+0.064}_{-0.055}$ $^{a}$ & 
$8.425^{+0.024}_{-0.028}$ $^{b}$ & Y& Y\\
20220307B & 0.248 $^{a}$ & $10.144^{+0.031}_{-0.036}$ $^{a}$ & $0.501^{+0.154}_{-0.170}$ $^{a}$ &
$8.621^{+0.012}_{-0.012}$ $^{b}$ & Y& Y\\
20221113A & 0.251 $^{a}$ & $9.476^{+0.043}_{-0.042}$ $^{a}$ & $-0.614^{+0.127}_{-0.144}$ $^{a}$ & 
$8.427^{+0.039}_{-0.050}$ $^{b}$ & Y& Y\\
20231123B & 0.262 $^{a}$ & $11.039^{+0.010}_{-0.009}$ $^{a}$ & $0.686^{+0.020}_{-0.019}$ $^{a}$ & 
$8.799^{+0.007}_{-0.007}$ $^{b}$ & Y& N\\
20230307A & 0.271 $^{a}$ & $10.761^{+0.026}_{-0.023}$ $^{a}$ & $-0.333^{+0.055}_{-0.059}$ $^{a}$ & 
$8.350^{+0.011}_{-0.011}$ $^{b}$ & Y& Y\\
20221116A & 0.276 $^{a}$ & $11.015^{+0.025}_{-0.021}$ $^{a}$ & $1.354^{+0.106}_{-0.136}$ $^{a}$ & 
$8.845^{+0.011}_{-0.011}$ $^{b}$ & Y& Y\\
20221012A & 0.285 $^{a}$ & $10.959^{+0.020}_{-0.016}$ $^{a}$ & $-0.739^{+0.182}_{-0.218}$ $^{a}$ & 
$8.674^{+0.007}_{-0.007}$ $^{b}$ & Y& N\\
20220506D & 0.300 $^{a}$ & $10.448^{+0.015}_{-0.028}$ $^{a}$ & $0.078^{+0.056}_{-0.066}$ $^{a}$ & 
$8.538^{+0.002}_{-0.002}$ $^{b}$ & Y& Y\\
20230501A & 0.301 $^{a}$ & $10.290^{+0.002}_{-0.021}$ $^{a}$ & $0.612^{+0.060}_{-0.050}$ $^{a}$ & 
$8.621^{+0.010}_{-0.011}$ $^{b}$ & Y& Y\\
20230626A & 0.327 $^{a}$ & $10.442^{+0.039}_{-0.045}$ $^{a}$ & $-0.007^{+0.126}_{-0.123}$ $^{a}$ & 
$8.634^{+0.006}_{-0.006}$ $^{b}$ & Y& Y\\
20220726A & 0.362 $^{a}$ & $10.184^{+0.039}_{-0.035}$ $^{a}$ & $-0.151^{+0.105}_{-0.115}$ $^{a}$ & 
$8.821^{+0.023}_{-0.027}$ $^{b}$ & Y& N\\
20220204A & 0.401 $^{a}$ & $9.699^{+0.036}_{-0.091}$ $^{a}$ & $0.634^{+0.122}_{-0.099}$ $^{a}$ & 
$8.568^{+0.002}_{-0.002}$ $^{b}$ & Y& Y\\
20230712A & 0.453 $^{a}$ & $11.126^{+0.015}_{-0.014}$ $^{a}$ & $1.476^{+0.010}_{-0.012}$ $^{a}$ & 
$8.683^{+0.013}_{-0.013}$ $^{b}$ & Y& N\\
20220310F & 0.478 $^{a}$ & $9.982^{+0.084}_{-0.059}$ $^{a}$ & $-0.013^{+0.122}_{-0.153}$ $^{a}$ & 
$8.615^{+0.025}_{-0.027}$ $^{b}$ & Y& Y\\
20221219A & 0.553 $^{a}$ & $10.207^{+0.033}_{-0.039}$ $^{a}$ & $0.251^{+0.055}_{-0.059}$ $^{a}$ &
$8.468^{+0.002}_{-0.002}$ $^{b}$ & N& Y\\
20220418A & 0.621 $^{a}$ & $10.258^{+0.024}_{-0.024}$ $^{a}$ & $0.118^{+0.048}_{-0.045}$ $^{a}$ &
$8.235^{+0.013}_{-0.014}$ $^{b}$ & N& Y\\
20221029A & 0.975 $^{a}$ & $11.031^{+0.0001}_{-0.0001}$ $^{a}$ & $1.653^{+9.7\times10^{-7}}_{-9.7\times10^{-7}}$ $^{a}$ & 
$8.042^{+0.044}_{-0.034}$ $^{b}$ & N& Y\\
20121102A & 0.193 $^{c}$ & $8.140^{+0.090}_{-0.100}$ $^{c}$ & $-1.301^{+0.146}_{-0.097}$ $^{c}$ & 
$>8.127^{+0.037}_{-0.039}$ $^{d}$ 
& N& Y\\
20180301A & 0.331 $^{c}$ & $9.640^{+0.110}_{-0.110}$ $^{c}$ & $0.281^{+0.126}_{-0.147}$ $^{c}$ & 
$8.566^{+0.014}_{-0.015}$ $^{d}$ & Y& Y\\
20180924B & 0.321 $^{c}$ & $10.390^{+0.020}_{-0.020}$ $^{c}$ & $-0.208^{+0.181}_{-0.213}$ $^{c}$ &
$8.710^{+0.008}_{-0.009}$ $^{d}$ & Y& N\\
20181112A & 0.475 $^{c}$ & $9.870^{+0.070}_{-0.070}$ $^{c}$ & $0.188^{+0.216}_{-0.238}$ $^{c}$ &
$8.655^{+0.022}_{-0.020}$ $^{d}$ & Y& N\\
20190102C & 0.291 $^{c}$ & $9.690^{+0.090}_{-0.110}$ $^{c}$ & $-0.398^{+0.249}_{-0.140}$ $^{c}$ & -- $^{d,\,e}$
% NII OIII UL 
& Y& Y\\
20190520B & 0.242 $^{c}$ & $9.080^{+0.080}_{-0.090}$ $^{c}$ & $-1.398^{+0.301}_{-0.301}$ $^{c}$ & 
$>8.094^{+0.040}_{-0.035}$ $^{d}$ % Hb NII UL
& N& N\\
20190608B & 0.118 $^{c}$ & $10.560^{+0.020}_{-0.020}$ $^{c}$ & $0.847^{+0.080}_{-0.078}$ $^{c}$ & 
$8.655^{+0.009}_{-0.009}$ $^{d}$ & Y& N\\
20190611B & 0.378 $^{c}$ & $9.570^{+0.120}_{-0.120}$ $^{c}$ & $-0.276^{+0.390}_{-0.293}$ $^{c}$ & 
$8.599^{+0.039}_{-0.033}$ $^{d}$ & Y& Y\\
20190714A & 0.236 $^{c}$ & $10.220^{+0.040}_{-0.040}$ $^{c}$ & $0.276^{+0.216}_{-0.208}$ $^{c}$ & 
$8.762^{+0.012}_{-0.012}$ $^{d}$ & Y& Y\\
20191001A & 0.234 $^{c}$ & $10.730^{+0.070}_{-0.080}$ $^{c}$ & $1.262^{+0.288}_{-0.292}$ $^{c}$ & 
$8.762^{+0.009}_{-0.010}$ $^{d}$ & Y& Y\\
20200430A & 0.161 $^{c}$ & $9.300^{+0.070}_{-0.100}$ $^{c}$ & $-0.959^{+0.189}_{-0.196}$ $^{c}$ &
$8.699^{+0.023}_{-0.016}$ $^{d}$ & N& Y\\
20200906A & 0.369 $^{c}$ & $10.370^{+0.050}_{-0.050}$ $^{c}$ & $0.693^{+0.231}_{-0.280}$ $^{c}$ & 
$8.607^{+0.011}_{-0.011}$ $^{d}$ & Y& Y\\
20201124A & 0.098 $^{c}$ & $10.220^{+0.050}_{-0.050}$ $^{c}$ & $0.435^{+0.206}_{-0.258}$ $^{c}$ & 
$8.758^{+0.008}_{-0.008}$ $^{d}$ & Y& Y\\
20210320C & 0.280 $^{c}$ & $10.370^{+0.050}_{-0.060}$ $^{c}$ & $0.545^{+0.229}_{-0.231}$ $^{c}$ &
$8.679^{+0.013}_{-0.014}$ $^{d}$ & Y& Y\\
20210410D & 0.141 $^{c}$ & $9.470^{+0.050}_{-0.050}$ $^{c}$ & $-1.523^{+0.301}_{-0.176}$ $^{c}$ &
$<8.791^{+0.020}_{-0.021}$ $^{d}$ 
& Y& Y\\
20210807D & 0.129 $^{c}$ & $10.970^{+0.020}_{-0.020}$ $^{c}$ & $-0.201^{+0.109}_{-0.137}$ $^{c}$ & 
$>8.614^{+0.007}_{-0.007}$ $^{d}$  
& Y& N\\
20211127I & 0.047 $^{c}$ & $9.480^{+0.060}_{-0.020}$ $^{c}$ & $1.554^{+0.012}_{-0.018}$ $^{c}$ &
$8.819^{+0.012}_{-0.009}$ $^{d}$ & Y& Y\\
20211203C & 0.344 $^{c}$ & $9.760^{+0.070}_{-0.090}$ $^{c}$ & $1.202^{+0.071}_{-0.090}$ $^{c}$ &
$8.678^{+0.011}_{-0.011}$ $^{d}$ & Y& Y\\
20220105A & 0.278 $^{c}$ & $10.010^{+0.050}_{-0.070}$ $^{c}$ & $-0.377^{+0.240}_{-0.262}$ $^{c}$ &
$8.697^{+0.016}_{-0.015}$ $^{d}$ & Y& Y\\
\end{longtable*}
\vspace{-1em}
{\footnotesize
\begin{flushleft}
\hspace{13em}
$^a$ adopted from \citet{sharma24} \\
\hspace{13em} 
$^b$ independently estimated from \citet{sharma24} emission line measurements \\
\hspace{13em} 
$^c$ adopted from \citet{gordon23} \\
\hspace{13em} 
$^d$ independently estimated from \citet{eftekhari23} emission line measurements \\
\hspace{13em} 
$^e$ upper limits on both O3 and N2, resulting in no constraint on O3N2 (see \S \ref{ss:metal})\\
\end{flushleft}
}

\section{Impact of Varying Model Assumptions on the Metallicity Offsets}
\label{s:testmetal}

\subsection{Mass and SFR Calibrations between Parametric and Non-Parametric SFH Methods}
\label{ss:leja22}

\begin{figure}
\centering
\includegraphics[width=10 cm]{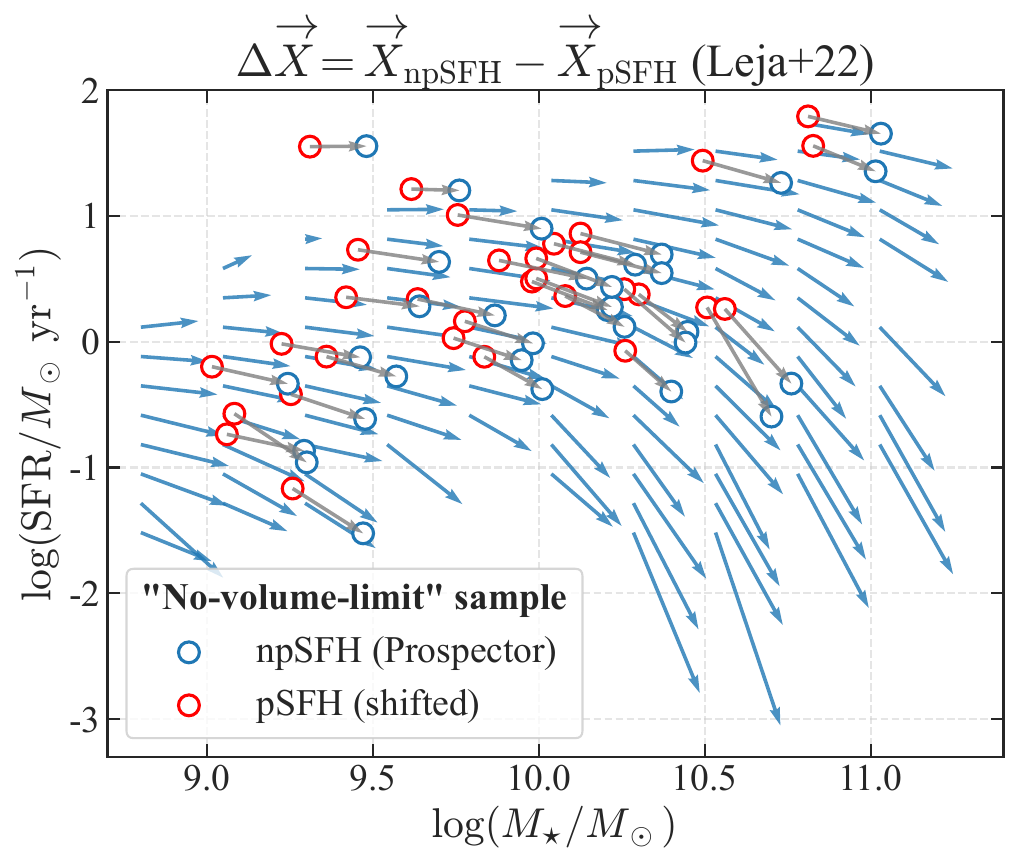}
\caption{Median shift in the inferred values of $\log M_\star$ and $\log{\rm SFR}$ between the Prospector non-parametric SFH (npSFH) and parametric SFH (pSFH) models for galaxies at $0.7 < z < 1.3$ (from Fig. 9 of \citealt{leja22}). Only cells with $N > 20$ galaxies are shown. The $x$- and $y$-axes are defined in the npSFH scale, and the blue vectors indicate the shift toward the pSFH scale ($\Delta \overrightarrow{X} = \overrightarrow{X}_{\rm npSFH} - \overrightarrow{X}_{\rm pSFH}$). The average offsets in $\log M_\star$ and $\log{\rm SFR}$ between the npSFH and pSFH models are $-0.2$ and $-0.3$ dex, respectively.
Blue open circles show the FRB host galaxies (“no-volume-limit” sample in Table \ref{tab:selection}) measured on the npSFH scale, while red circles indicate their positions after conversion to the pSFH scale. Gray arrows represent the corresponding shift vectors for individual hosts, matched to the nearest available vector in the grid.}
\label{fig:fig9leja22}
\end{figure}

\begin{figure*}
\centering
\includegraphics[width=17cm]{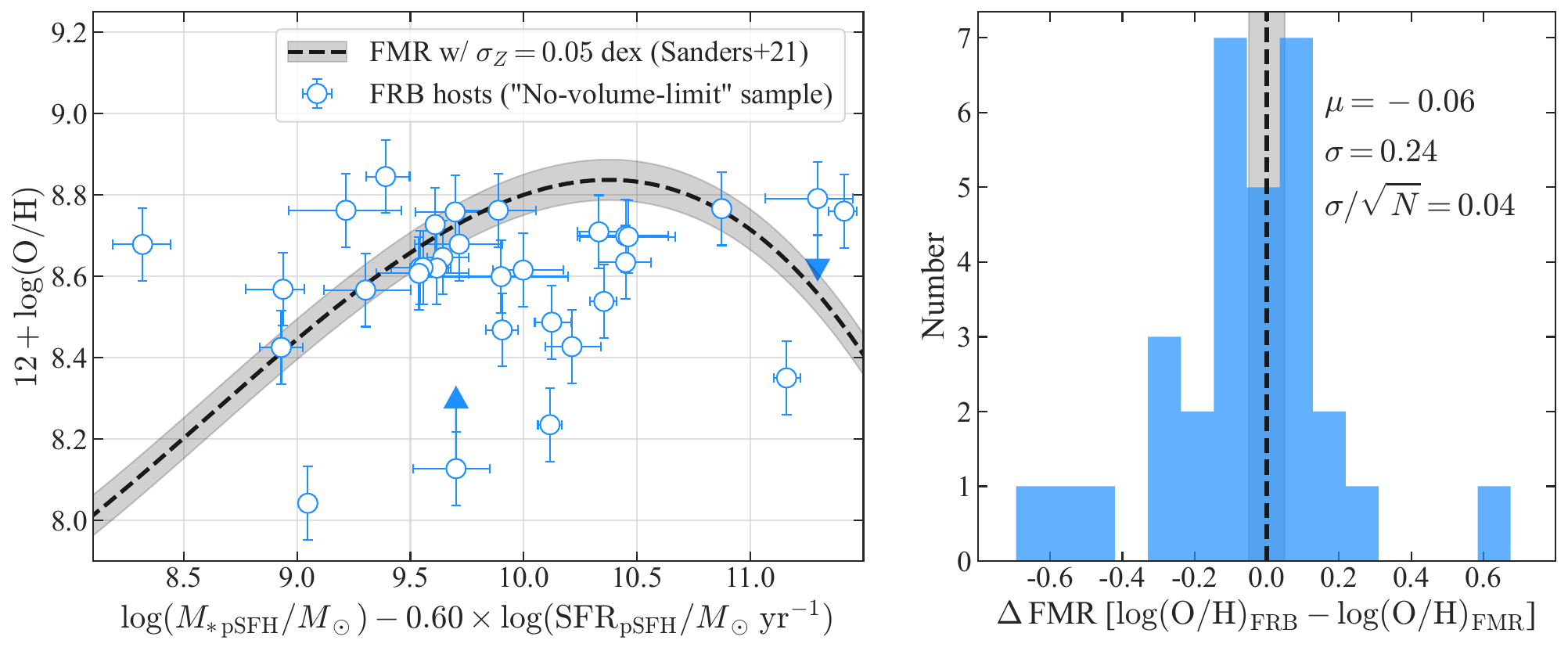}
\caption{Same as Figure \ref{fig:fmr}, which applied only the mass correction, but here both mass and SFR corrections have been applied. Note that unlike Figure \ref{fig:fmr}, the $x$-axis is shown in the pSFH scale rather than the npSFH scale.}
\label{fig:fmr_msfr_shifted}
\end{figure*}

Although we applied a uniform 0.2 dex mass correction throughout the main body of the paper, we have now explicitly tested the effect of applying both mass and SFR corrections depending on the galaxy’s location in the mass–SFR plane. This is motivated by the fact that non-parametric SFH (npSFH) SFRs derived with Prospector may be systematically lower by up to $\sim$1 dex compared to parametric SFH (pSFH) measurements (e.g., \citealt{leja19}). To this end, we adopt the shifts between npSFH and pSFH from Fig. 9 of \citet{leja22} for galaxies at $0.7 < z < 1.3$. Strictly speaking, \citet{leja22} use UV+IR–based SFRs as a proxy for pSFH SFRs, and we likewise treat them as an approximate pSFH proxy for simplicity (see \S \ref{ss:model_metal}). Also, we are aware that our FRB host galaxy sample (a non–volume-limited sample, with the majority at $z < 0.5$) has smaller redshifts than this, but this is the only available sample tested in the literature. We then convert the masses and SFRs of the FRB hosts from the npSFH scale to the pSFH scale using these shift vectors, depending on their position in the mass–SFR plane (Figure \ref{fig:fig9leja22}). Next, we compare the resulting pSFH-scale values with the FMR measured on the pSFH scale in a consistent manner (Figure \ref{fig:fmr_msfr_shifted}). The resulting metallicity offsets are reduced to $-0.06 \pm 0.04$~dex ($1.5\sigma$), compared to the previous $-0.09 \pm 0.04$~dex ($2.3\sigma$) obtained using only a uniform mass shift of $0.2$~dex. Thus, while applying the mass and SFR shifts decreases the apparent offset from the FMR, the FRB hosts still remain systematically below the relation. This demonstrates that the offset cannot be fully explained by the lack of an SFR correction, even if the applied correction is valid across different redshift ranges.

\subsection{Adopting An Alternative FMR}
\label{ss:curti20}

To test the potential impact of adopting alternative values of $\alpha$ from the literature, we performed an additional analysis using $\alpha = 0.55$ from \citet{curti20} (``Global sample'' in their Table~A1). They parameterized the FMR with the following functional form:
\beq
12 + \log(\mathrm{O/H}) = Z_0 - \frac{\gamma}{\beta} \log\left[ 1 + \left( \frac{10^{\mu_{0.55}}}{10^{\mu_0}} \right)^{-\beta} \right],
\eeq
where $Z_0=8.780$, $\gamma=0.30$, $\beta=2.4$, and $\mu_0=10.14$. The \citet{curti20} relation exhibits an almost linear slope for $\mu_{0.55} \lesssim \mu_0$, flattening above $\mu_0$. This is in contrast to the \citet{sanders21} relation, which is described by a cubic function of $\mu_{0.60}$ (Eq.~\ref{eq:fmr}). The resulting metallicity offsets are reduced to $-0.03 \pm 0.04$~dex ($0.8\sigma$), compared to the previous $-0.09 \pm 0.04$~dex ($2.3\sigma$).
This suggests that the choice of FMR can significantly affect the measured offsets. However, when both stellar-mass and SFR corrections are applied (as in Section~\ref{sss:FMR_correction} and Appendix~\ref{ss:leja22}), the offset becomes $-0.07 \pm 0.04$~dex (1.8$\sigma$), indicating that a residual offset may still persist.

\begin{figure*}
\centering
\includegraphics[width=17cm]{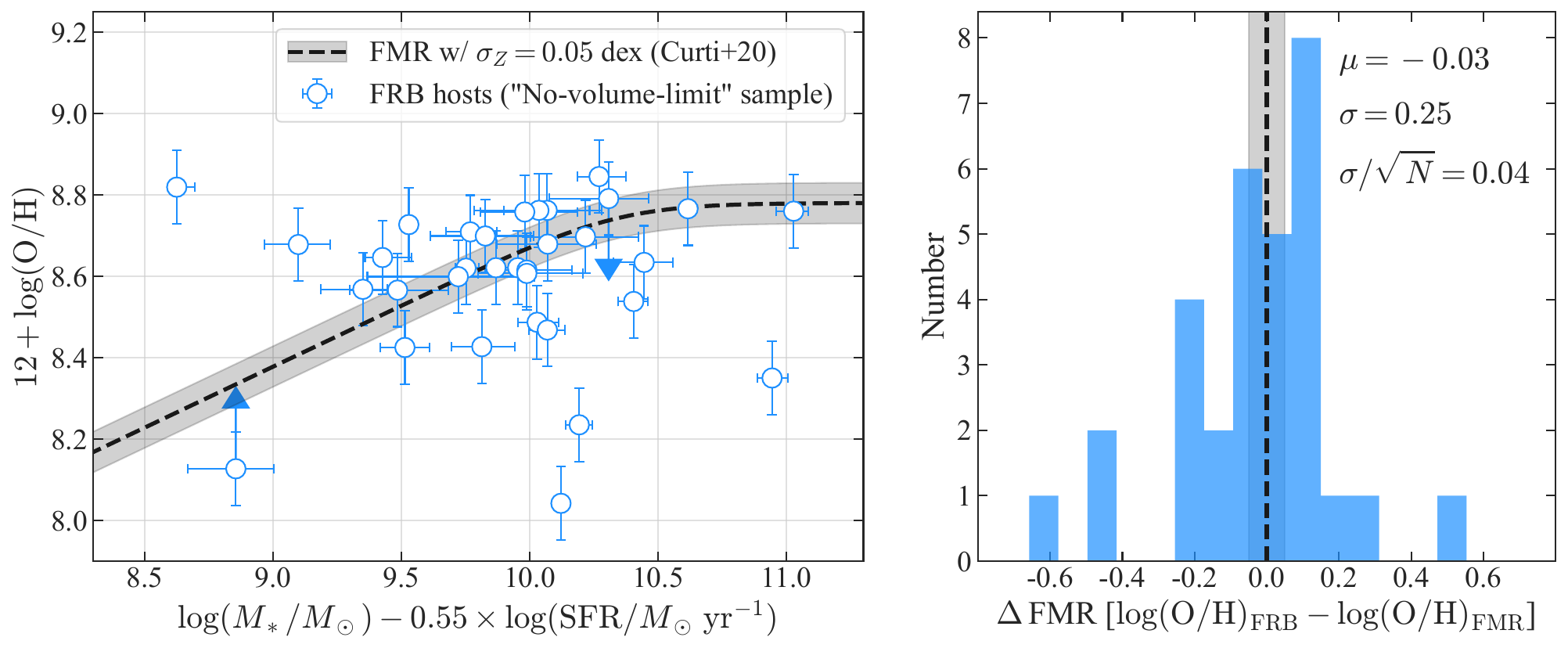}
\caption{Same as Figure~\ref{fig:fmr}, which applies the FMR from \citet{sanders21} with $\alpha=0.60$, but here we adopt the alternative FMR from \citet{curti20} with $\alpha=0.55$ (see text).}
\label{fig:fmr_alpha055}
\end{figure*}


%\bibliography{frb_metal}{}
\bibliographystyle{aasjournal}

\end{document}